\documentclass[10pt,conference]{IEEEtran}

\usepackage{graphicx}
\usepackage{multirow}
\usepackage{booktabs}
\usepackage{colortbl}
\usepackage{xcolor}
\usepackage{multicol}
\usepackage{rotating}
\usepackage{tabularx}
\usepackage{bigstrut}
\usepackage{amssymb}
\usepackage{rotating}
\usepackage{upgreek}
\usepackage{balance}
\usepackage[bookmarks=false]{hyperref}

\makeatletter
\newcommand{\ygg@basicalert}[2]{\fbox{\bfseries\sffamily\scriptsize#1}{\sf\small$\blacktriangleright$\textit{#2}$\blacktriangleleft$}}
\newcommand{\BONITA}[1]{\ygg@basicalert{BONITA}{#1}}
\newcommand{\ALL}[1]{\ygg@basicalert{ALL}{#1}}
\newcommand{\SUGGESTION}[1]{\ygg@basicalert{ALL}{#1}}
\newcommand{\NAHLA}[1]{\ygg@basicalert{NAHLA}{#1}}
\newcommand{\JONATHAN}[1]{\ygg@basicalert{NAHLA}{#1}}

\usepackage[sort,compress]{cite}

\usepackage[numbers, sort&compress]{natbib}

\ifCLASSINFOpdf
\else
\fi
\usepackage{algorithmic}

\usepackage{array}
\begin{document}
%

\title{Developer Reading Behavior While Summarizing Java Methods: Size and Context Matters}




%
\author{\IEEEauthorblockN{Nahla J. Abid\IEEEauthorrefmark{1},
Bonita Sharif\IEEEauthorrefmark{2},
Natalia Dragan\IEEEauthorrefmark{3}, 
Hend Alrasheed\IEEEauthorrefmark{4} and
Jonathan I. Maletic\IEEEauthorrefmark{3}}
\IEEEauthorblockA{\IEEEauthorrefmark{1}Department of Computer Science, Taibah University, Madinah, Kingdom of Saudi Arabia}
\IEEEauthorblockA{\IEEEauthorrefmark{2}Department of Computer Science and Engineering, University of Nebraska-Lincoln, Lincoln, Nebraska, USA 68588}
\IEEEauthorblockA{\IEEEauthorrefmark{4}Department of Information Technology, King Saud University, Riyadh, Kingdom of Saudi Arabia}
\IEEEauthorblockA{\IEEEauthorrefmark{3}Department of Computer Science, Kent State University, Kent, Ohio, USA 44242\\Emails: nabd@taibahu.edu.sa, bsharif@unl.edu, ndragan@kent.edu, halrasheed@ksu.edu.sa, jmaletic@kent.edu}
}


\maketitle
\thispagestyle{plain}
\pagestyle{plain}

\begin{abstract}
An eye-tracking study of 18 developers reading and summarizing Java methods is presented. The developers provide a written summary for methods assigned to them.  In total, 63 methods are used from five different systems.  Previous studies on this topic use only short methods presented in isolation usually as images.  
In contrast, this work presents the study in the Eclipse IDE allowing access to all the source code in the system.  The developer can navigate via scrolling and switching files while writing the summary.  New eye-tracking infrastructure allows for this improvement in the study environment.  Data collected includes eye gazes on source code, written summaries, and time to complete each summary.  
Unlike prior work that concluded developers focus on the signature the most, these results indicate that they tend to focus on the method body more than the signature.  Moreover, both experts and novices tend to revisit control flow terms rather than reading them for a long period.  They also spend a significant amount of gaze time and have higher gaze visits when they read call terms.  
Experts tend to revisit the body of the method significantly more frequently than its signature as the size of the method increases.
Moreover, experts tend to write their summaries from source code lines that they read the most.
\end{abstract}


%

\IEEEpeerreviewmaketitle

\section{Introduction}
\label{sec:intro}

Source code  reading and comprehension is an essential and time-consuming task that programmers perform during software maintenance \cite{ko2006, LaToza2006}.  Natural language documentation and code summarizations are found to be critical to improve code comprehension \cite{Fluri2007}. In fact, expert and novice programmers tend to read comments more than the source code during comprehension activities \cite{Crosby1990}.  Unfortunately, comments are oftentimes incomplete \cite{deSouza2005} or outdated due to changes in code during maintenance.

One way to overcome this problem is to automatically generate summaries directly from source code. Several approaches are proposed to generate automatic summaries using Natural Language Processing (NLP) \cite{Sridhara2010,McBurney2014}, text retrieval \cite{Haiduc2010, Eddy2013}, and static analysis \cite{Moreno2013,Abid2015}. In order to further improve source code summarization techniques \cite{Sridhara2010, McBurney2014, Haiduc2010, Abid2015, Sridhara2011a, Sridhara2011b,Badihi2017, wan2018improving,Moreno2017}, Rodeghero et al. conducted an eye-tracking study \cite{Rodeghero2014} to determine the statements and terms (i.e., identifier names) that programmers view as important when they summarize a method. Their main results indicate that programmers tend to look at method signatures the most. They use this information to give more weight to terms that developers look at more often during summarization (i.e., signature is given more weight compared to call statements) as heuristics for their automated summarization approach that was adapted from prior work \cite{Haiduc2010}.

Unfortunately the prior eye-tracking study by Rodeghero et al. \cite{Rodeghero2014} is limited in that the study could not be conducted in a realistic working environment.  This is a common limitation when using eye-tracking equipment.  The work presented here focuses on addressing the two main limitations in the previous study.  The first limitation is that the maximum length of methods included in their study is only 22 lines. That is, a single fixed window of source code that must appear all at once on the screen with \textit{no} support for scrolling.  The second limitation is that methods are presented in complete isolation with no ability to look at other methods or related code.  

We overcome these two limitations by utilizing a new eye-tracking infrastructure call iTrace \citep{Shaffer2015,Guarnera18iTrace} that is developed specifically for conducting eye-tracking studies within a software Integrated Development Environment (e.g., MS Visual Studio, Eclipse).   iTrace supports the implicit tracking of eye movements in the presence of scrolling and switching between files within the IDE.  Hence we are able to conduct our study using any sized methods and participants have access to all of the files and the entire code base for the system being studied.  That is, the study is conducted in the same environment programmers use daily thus avoiding bias in the experimental setup and allowing participants to read code as they normally would. The contributions of this work are as follows: 
\begin{itemize}
  \item An eye-tracking study of 18 developers reading and summarizing Java methods in the Eclipse integrated development environment. Unlike the prior study that concluded developers mostly focus on the signature \cite{Rodeghero2014}, the results from our study indicate that developers tend to focus on the method body more than the signature.

  \item An analysis of programmers' gaze time and number of visits on source code entities with respect to the method size. For experts, we found that the larger the method, the less likely its signature is focused on. However, as the control flow increases in complexity, it is revisited more. Novices do not seem to be affected by the method size.
  
  \item An analysis of locations outside the method scope that developers target during method summarization. The results reveal that experts and novices read other locations besides the assigned methods (e.g., data members and class declarations) for about 70\% of the tasks.

\item A comparison of the lines used by experts to write their summaries to the lines with the longest gaze time. We found strong evidence that developers tend to write their summaries from source-code lines that they read the most (longest gaze time). 
Therefore, we conclude that gaze time can substantially predict lines that are important for summarizing a method \cite{Sridhara2010, McBurney2014, Abid2015, Sridhara2011a, Rodeghero2014}.

\end{itemize}
  
  
An overarching goal of this paper is to investigate ways to improve existing summarization approaches via findings from the eye-tracking study presented. This is done by examining what developers read during summarization in addition to what they include in their summaries. This is one of the first studies done in a realistic manner similar to how a developer would work in the field, i.e., the source code is presented in an IDE instead of a fixed image presented in isolation. This work also impacts overall tool development for software engineering researchers and practitioners and learning about differences between experts and novices. 

We present related work on eye tracking in software engineering and code summarization in Section \ref{sec:related}. Section \ref{sec:rqs} outlines our research questions. Section \ref{sec:design} describes the design of the eye tracking study. Analyses, results, and threats are presented in Sections \ref{sec:preprocess} through \ref{sec:threats} followed by implications in Section \ref{sec:discussion}. We end with final remarks in the conclusion along with future work.

\section{Related Work}
\label{sec:related}

We focus mainly on relevant eye tracking studies done in code comprehension.  We also discuss relevant work on code summarization given our work impacts that topic.  

\textbf{Eye Tracking Studies:} Eye-tracking technology is being used in software engineering \cite{Sharafi2015,Sharif2017,Sharafi2015d,Obaidellah2018,Barik2017,Bednarik2006} to study how programmers read \cite{Crosby1990,Bednarik2008,Busjahn2015,Rodeghero2015ESEM,Jbara2017,Lee2017}, review \cite{Uwano2006,Sharif2012,Begel2018}, and summarize \cite{Rodeghero2014,Rodeghero2015ESEM,Rodeghero2015TSE} source code.
Crosby et al. \cite{Crosby1990} conducted an early eye-tracking study of high and low experience programmers reading a binary search algorithm.   Both high and low experience programmers needed a large amount of fixations in most areas of source code than participants that read simple natural language text. The programmers tend to alternate between source code and comments rather than sequentially reading the entire document.

Uwano et al. \cite{Uwano2006} observed that programmers tend to first read through the entire code snippet, and then focus on some parts. Furthermore, longer time spent thoroughly reading the code increases the efficiency of finding the defect in the code. This correlation was later confirmed by Sharif et al. \cite{Sharif2012} stating that the scan time plays an important role in defect detection time and visual effort required to review source code.  Moreover, experts tend to focus on defects more than novices \cite{Sharif2012}. Busjahn et al. \cite{Busjahn2015} found that experts read code less linearly than novices did. Bednarik and Tukiainen concluded that low-experience programmers repeatedly fixated on the same code sections, while experienced programmers target the output of the code, such as evaluation expressions \cite{Bednarik2006,Bednarik2008}.


Kevic et al. used iTrace to conduct an eye tracking study on three bug fixing tasks.  They found that developers focus on small parts of methods that are often related to data flow.  When it comes to switches between methods, they found developers rarely follow call graph links and mostly switch to the elements in close proximity \cite{Kevic2015,Kevic2017}. 

Rodeghero et al. conduct an eye-tracking study on isolated methods to determine statements and terms that programmers view as important when they summarize them \cite{Rodeghero2014}.  They conclude that programmers consider method signatures as the most important section of code followed by call terms then control flow terms.  Based on these results, a modified weighted scheme for the Vector Space Model (VSM), an information retrieval method is implemented. The frequency of terms in VSM is replaced by the term positions. Therefore, terms located in the method signature have a higher weight than method invocation and method invocations have higher weight than control flow statements.  When the term-based summarization generated using the original VSM and their improved version is compared to human term-summaries, the improved version outperformed the original VSM. In an extended version \cite{Rodeghero2015TSE} of their initial work, they concluded that terms developers used in their method summaries generally have higher tf/idf scores than terms with high gaze times.
In addition, they concluded that “longer viewed” terms are mostly long terms (number of characters) and they are less likely to appear in developers’ summaries.  The same study \cite{Rodeghero2014} was further analyzed in \cite{Rodeghero2015ESEM} to find reading patterns of programmers during summarization.  
On average, they find that programmers read from top-to-bottom about 49\% of the time.  Finally, they conclude that all of the programmers followed nearly identical eye patterns in overall reading.

The study we present in this paper is similar to the one performed by Rodeghero et al. \cite{Rodeghero2014} with several differences. One of our goals was to study if we get the same results after mitigating limitations related to methods size and the study environment as described earlier. In addition, our study compares experts and novices eye movements, and analyzes them with respect to methods’ sizes.

\textbf{Code Summarization:} With regards to code summarization approaches, Sridhara et al. propose techniques to automatically generate natural language comments for Java methods \cite{Sridhara2010}, sequences of statements \cite{Sridhara2011b}, and formal parameters \cite{Sridhara2011a} using NLP. 
Furthermore, Wang et al. propose a model that defines the high-level action of loops by analyzing linguistic and structure clues \cite{Wang2015}.  They also presented an approach to automatically generate a natural language summary of object-oriented action units \cite{Wang2017}. 

Moreno et al. \cite{Moreno2013} use method stereotypes \cite{Dragan2006} and class stereotypes \cite{Dragan2010} to generate natural language summaries for Java classes.  Abid et al. \cite{Abid2015} \cite{Abid2017} use method stereotypes to generate a standard summary for C++ methods via static analysis. McBurney and McMillan propose generating documentation summaries for Java methods using the call graph \cite{McBurney2014}. Furthermore, they propose an approach to evaluate a summary using textual similarity of that summary to the source code \cite{Mcburney2016}. Haiduc et al. \cite{Haiduc2010} investigate the suitability of several text summarization techniques  to automatically generate term-based summaries for methods and classes.  This was further extended by Eddy et al. \cite{Eddy2013} using a new technique named Hierarchical PAM. 



\section{Research Questions}
\label{sec:rqs}

The \textit{first} goal of this study is to understand what terms developers focus on when they summarize Java source code using a realistic integrated development environment setting.  The analysis considers three source-code locations namely, method signatures, method calls, and control flow statements \cite{Rodeghero2014,Rodeghero2015TSE,Rodeghero2015ESEM}. These locations were chosen so that we can make comparisons to the Rodeghero study \cite{Rodeghero2014}. A term refers to an identifier in a specific location, e.g., call terms are identifiers in a call. 
In addition to the above, our work expands the analysis in several different directions based on the level of expertise of programmers and size (length) of methods. 
The \textit{second} goal is to investigate how experts read source-code lines during summarization tasks and how this information can be used to improve source-code summarization and more broadly documentation. The \textit{third} goal of our work considers locations outside the method scope that programmers might find important such as the class name, data member declarations, and other related methods. Such scope analysis could not be done in the Rodeghero study as the methods were presented in isolation \cite{Rodeghero2014}. To address the above three goals, we pose the following research questions.

\begin{itemize}
\item [RQ1]	Considering all methods summarized, to what extent do experts and novices focus on a method's signature, method's body, call invocations in a method, and method's control flow?

\item [RQ2]	Does the size (length) of the method (small, medium, large) have an impact on what experts and novices look at most during the summarization task?

\item [RQ3]	What source code elements (if any) does a programmer look at outside the scope of the assigned method to summarize? How do they use context around the method in the summarization task?


\item [RQ4] What source code lines do experts read and use in writing summaries?

\end{itemize}

The first research question will help us compare our results with the Rodeghero study albeit in a more realistic setting. The second research question digs deeper into the effect method size has on reading for summarization. The third research question focuses on context around the method, i.e., what else besides the method being summarized are looked at in the source code (note that unlike the Rodeghero study, in this study the entire source code was provided to the participants with eye tracking data recorded at the line and term level), Finally, the fourth research question compares what was looked at with what was written in the summaries.  This analysis was restricted to just experts as they write accurate summaries.

\section{Eye Tracking Study - Experimental Design }
\label{sec:design}
This section describes the process of designing and performing the eye-tracking study for method summarization tasks. The study material including tasks, processed eye-tracking data and the statistical analysis is provided in the replication package at 
\url{https://doi.org/10.5281/zenodo.2550768}.

\subsection{Study Participants}
The study is performed by 18 developers, a typical sample size for eye-tracking studies. They are 5 experts and 13 novices. Two of the experts are industry professionals working at a local firm and three are PhD students at a local university. One industry expert had between two to five years of programming experience and the other four had greater than 5 years of programming experience. We consider the PhD students as experts because they are heavily involved in coding for open source projects. Novices are undergraduate and graduate students 
with one to five years of programming experience, most with about a year of experience.

\begin{table}[tp]
\caption{Java Systems Used in the Study}
\label{tab:systems}
\resizebox{\columnwidth}{!}{%
\centering
\begin{tabular}{ll  rrr}
\hline
\textbf{System version} & \textbf{Domain} & \textbf{Total methods} & \textbf{Selected methods} & \textbf{Classes involved} \\
\hline
ArgoUML 0.30.2 & UML diagramming tool & 14,635 & 15 & 2,673 \\
MegaMek 0.36.0 & Computer game & 12,490 & 15 & 2,308 \\
Siena 1.0.0 & Database library & 4,116 & 12 & 297 \\
sweetHome3d4.1 & Interior design application & 6,084 & 12 & 1,757 \\
aTunes 3.1.0 & Audio player & 9,579 & 9 & 215 \\
\hline
\textbf{Total} & \textbf{} & \textbf{46,904} & \textbf{63} & \textbf{7,250} \\
\hline
\end{tabular}%
}
\end{table}

\subsection{Study Systems and Method Selection}
The study includes 63 methods from five open source Java systems randomly selected from different domains (see Table \ref{tab:systems}). While the selection of methods chosen to be summarized is random, we maintain two conditions. First, we eliminate trivial methods such as setters, getters, and empty methods. Second, the largest method (excluding blank lines) chosen was set to 80 LOC. This was done to avoid excessive fatigue during tasks. Based on the above criteria, a total of 63 methods were selected from the systems shown in Table \ref{tab:systems}.  
Participants were randomly given a set of 23 methods selected from the 63 methods to summarize. 
On average, participants summarized 15 of the 23 methods. 

The size of short, medium and long methods range between 9-22 LOC, 23-39 LOC, and 40-80 LOC respectively.  A line of code is counted if and only if it is not empty and is not a comment.  We used this split to maintain balanced number of methods in each size category.  Methods in Rodeghero et al.'s study \cite{Rodeghero2014} fall in the first category.  Therefore, methods larger than 22 LOC are analyzed separately to study the impact of summarizing larger methods. Another difference with the Rodeghero study is that we did not modify the code. In the Rodeghero study, method text was reformatted to make it fit on one line on the screen since they were shown in isolation. In our study, we left the methods as they appear in the code and did not reformat lines.  


\subsection{Task}
The participants were told that their main task was to read the assigned methods and write a summary for the method. They were also told that they could navigate the codebase if they needed to. The entire study was conducted inside the Eclipse environment using the Eclipse plugin iTrace \cite{Shaffer2015}. iTrace is able to collect eye tracking data of where a developer is looking and map it on the fly to source code elements looked at even in the presence of file scrolling and file switching. Code folding was disabled for all participants to avoid any confounding factors. The Eclipse environment was setup with all the projects and assigned methods open in various tabs. No web browser was used in this study.  The participants were also able to view the method while they were writing the summary. We extended iTrace for this study to include on-the-fly mapping to more source code constructs such as conditional and looping structures in addition to what was collected in an earlier study (method calls, signatures, and definition and use of variables) \cite{Kevic2015}.

\subsection{Eye Tracking Apparatus and Measures}
A Tobii eye tracker (X60) was used to collect gaze data within the iTrace \cite{Shaffer2015} environment. The eye tracker generates 60 raw gaze samples per second. The eye gaze is then passed through a fixation filter to generate fixations. A fixation is the stabilization of the eyes on some object of interest for a certain duration \cite{Rayner1998}. As in \cite{Rodeghero2014}, two types of eye-movement data are used: number of fixations and their durations (gaze time). 
We define gaze time as the total number of milliseconds spent on a region of interest (ROI) such as a call or keyword. A fixation filter \cite{Olsson2007} is set to count fixations that are more than 100 milliseconds (same as Rodeghero and vendor recommended). Throughout the analysis, we refer to the number of fixations as the number of visits on particular regions of interest.

\subsection{Study Procedure and Instrumentation}
We first obtained IRB approval for the study. On the day of the study, the participants first signed an informed consent form and filled out a background questionnaire. Next, they were given a short description on how the study would be conducted. They were also given a 1-page overview of each system and a hard copy of all the method names {(body not included)} to be summarized in case they needed to refer to it. Before beginning the actual study, they were given examples of three method summaries (by the original developers) taken from the same systems shown in Table \ref{tab:systems}. The sample methods are not part of the subsystems used in the study to avoid any learning effects. These examples included the method's source code  along with the summary for the methods.  This is a necessary step as it was important for the participants to understand what they were expected to do during a summarization task. We encouraged them to use their own words to summarize the methods. 
The study took between 45-90 mins.

The participants were seated in front of a 24-inch LCD monitor. They were not required to run or build the software systems. Before the study began, we calibrated each participant using a 9-point calibration.  This is a necessary step for the eye tracker to correctly capture data for each participant. A moderator was responsible to start and stop eye tracking within iTrace for each method to be summarized. This was done to have consistency in the starting and stopping of sessions across all participants. The participants wrote their summary in a text file present at the bottom of the Eclipse IDE.  They were able to see the code while writing the summary.  The font size was set to 14 points in the Eclipse editor. Gazes on the text files were also collected. At the end of each task, we had a time-stamped session of line-level gaze data on the source code and summary file that we used to answer our research questions.

\section{Pre-Processing the Data}
\label{sec:preprocess}
Before running statistical analysis on the data, we weeded out samples that were invalid. In addition, we identified the location (call, signature, control flow) a term belongs. 


\textit{Data Cleaning:} Some invalid samples include writing a narration of the code word for word (e.g.,"The method uses an if statement to determine if values are true"). Other cases were because participants were unable to locate the assigned method or were unable to understand the assigned method. In order to catch such cases, three of the authors reviewed the 257 summaries written by developers to judge their suitability for further analysis.  Each human summary is reviewed by two reviewers.  When both reviewers agree about the inclusion of the summary and its corresponding gaze data for analysis, it is kept otherwise it is discarded. The computed inter-rater reliability using Cohen's kappa ($\kappa$) is 0.92. There was high agreement between reviewers on what to discard. 
At the end of this step, 44 summaries (6 expert summaries and 38 novice summaries) are discarded.  


\textit{Term Location Identification.} The eye gaze data are processed to identify whether the term looked at corresponds to a signature, call, or control flow term. Information from the eye-tracking tool, iTrace \cite{Shaffer2015}, includes information about the line number, the term name and the type of the term (method, variable, conditional statements, or other types).  In some cases, the exact term that a participant looked at was not identified at the time of this experiment but the line number was. Since the code was not edited during the study, we use the line number to determine in a post processing step the nature of the location.  
First, for all methods used in the study, the signature lines are extracted from srcML \cite{Collard2011}, an XML representation of the code.  Next, if the line of an eye-tracking record matches the signature line in the list, the record is considered a signature.
The control flow statements are also identified at the line level because 
all of them stand on their own and do not have multiple statements listed on the same line, i.e., the body of an if statement appears on another line.  This simplifies the line-level analysis. 

Call terms need to be handled differently as the line containing a call might include other terms that are not part of the call.  We define call terms to include the method call name and parameters \cite{Rodeghero2014}. For example, in the statement, \texttt{boolean saved = saveFile(toFile)}, call terms include \texttt{saveFile} and \texttt{toFile}. 
The possibility of identifying call terms depends on the way the call is written.  All calls of our methods are manually examined to identify call cases. There were 49,366 total number of eye-tracking gazes inside all methods for all participants. 18,602 of those records are successfully identified as calls (38\%). 28,275 were non-call records (57\%).  The number of eye-tracking records that cannot be categorized as calls or non-calls is 2,488 (5\%). We exclude this 5\% from the analysis of calls vs. non-calls in RQ1 and RQ2. RQ3 and RQ4 are not affected by this identification.

\section{Experimental Results}
\label{sec:results}
We present our findings on each research question along with a comparison to the Rodeghero study when appropriate. Because our data is not normally distributed, the Wilcoxon non-parametric test is used. Additionally, since the same data is used to test multiple hypothesis, we use Bonferroni p-value correction. We perform three tests (signature, call, and control-flow terms) on each set of samples. Therefore, the new p-value to test against is (0.05/3)= 0.016. Anything before this value is considered significant. 
Effect size is given using Cohen's d.

\subsection{RQ1 Results: Replicating the Rodeghero study analysis }
 
Methods are read differently if presented in isolation vs. presented in a realistic environment that developers usually work in \cite{Kevic2015}.  To this end, we perform the same analysis done by Rodeghero et al. to study the impact of the differences. 


\textbf{Adjusted Metrics.} Number of terms in the  signature, call and control flow vary based on the size of the method. Therefore, it is necessary to adjust the total gaze time according to the number of terms in each method, which is the same adjustment performed by Rodeghero et al. \cite{Rodeghero2014}. Let's assume one reads the body and signature of a method for $60\%$ and $40\%$ of the time respectively. Consider the body and signature to be $80\%$ and $20\%$ in terms of overall size of identifiers. This does not mean that the body is more important. Longer time in this case is due to size of the body. If the body is $80\%$ of the overall size; the body is read $60/80= .75$ and the signature is read $40/20=2$. 2 is greater than .75; therefore the participant gives signature higher attention. Then, these two adjusted values are compared using the Wilcoxon non-parametric test for all collected samples. A sample is a result from one developer summarizing one method.

The same approach is used to compute the adjusted number of visits for the signature. The approach to compute the adjusted gaze time and number of visits for signature and non-signature terms is also applied to call terms vs. non-call terms, and control flow terms vs. non-control flow terms. A total of 26 methods are summarized by five experts and 40 methods are summarized by 13 novices. The analysis of the eye-movements of experts and novices are discussed for method signature, calls, and control flow terms. The adjusted gaze time and number of visits are used in hypothesis testing for RQ1 and RQ2. 

\textit{Method Signature:} 
We found no evidence that experts read method signatures more than the method body. On average, experts spent 9\% of their gaze time reading signatures, while the signatures averaged 10\% of the methods.  Similarly, the average gaze time of novices reading signatures is 10\% and signatures are 11\% of the methods. We computed the adjusted gaze time and number of visits of the signature and the body and pose four hypotheses ($H_1$, $H_2$, $H_3$, and $H_4$) as follows. 

\textit{$H_n$: For [experts / novices], the difference between the adjusted [gaze time / visit] metric for method signature and method body is not statistically significant.}

Based on computed $Z$ and $p$ values (see Table \ref{tab:mainresult}), we reject $H_2$ (medium effect size of $0.30$), $H_3$ (small effect size of $0.20$), and $H_4$ (medium effect size of $0.25$). This suggests that novices read/visit method bodies more than the method signatures.  We cannot reject $H_1$ which means that there is no statistical difference in the time spent by experts reading the signature and the body of a method.  However, experts revisit the method bodies more frequently than the method signatures.  When data of both experts and novices are combined, the results of Wilcoxon test for gaze time and number of visits is ($Z$= -3.93, $p$ \textless 0.001 with a small effect size of $0.19$) and ($Z$= -5.37, $p$ \textless 0.001 with a medium effect size of $0.26$), respectively. This result indicates that developers read method bodies more heavily than the method signatures when methods are presented in context.  This behavior was previously observed by Kevic et al. \cite{Kevic2015}.

\begin{table}[]
\centering
\caption{Wilcoxon Test Results of Experts and Novices. n is number of samples. A sample is a result from one developer and one method. T is the sum of ranks.}
\label{tab:mainresult}
\resizebox{\columnwidth}{!}{%
\begin{tabular}{l  l  l  l  r r r r}
\hline
 \textbf{} & \textbf{H} & \textbf{Metric} & \textbf{Location} & \textbf{$n$} & \textbf{$T$} & \textbf{$Z$} & \textbf{$p$}  \\
 \hline
\multirow{4}{*}{\textbf{Experts}} & \multirow{2}{*}{$H_1$} & \multirow{2}{*}{Gaze} & Signature & 69 & 905 & -1.809 & 0.064 \\
 &  &  & Non-Sig & 69 & 1510 &  &  \\
 & \multirow{2}{*}{$H_2$} & \multirow{2}{*}{Visit} & Signature & 69 & 623 & -3.49 & \textless{}.0001* \\
 &  &  & Non-Sig & 69 & 1792 &  &  \\
 \hline 
\multirow{4}{*}{\textbf{Novices}} & \multirow{2}{*}{$H_3$} & \multirow{2}{*}{Gaze} & Signature & 144 & 3533 & -3.36 & .001* \\
 &  &  & Non-Sig & 144 & 6907 &  &  \\
 & \multirow{2}{*}{$H_4$} & \multirow{2}{*}{Visit} & Signature & 144 & 3101 & -4.22 & \textless{}.0001* \\
 &  &  & Non-Sig & 144 & 7339 &  &  \\
 \hline 
\multirow{4}{*}{\textbf{Experts}} & \multirow{2}{*}{$H_5$} & \multirow{2}{*}{Gaze} & Call & 69 & 1655 & -2.67 & \textless{}.007* \\
 &  &  & Non-call & 69 & 760 &  &  \\
 & \multirow{2}{*}{$H_6$} & \multirow{2}{*}{Visit} & Call & 69 & 1554 & -2.07 & .03* \\
 &  &  & Non-call & 69 & 861 &  &  \\
 \hline 
\multirow{4}{*}{\textbf{Novices}} & \multirow{2}{*}{$H_7$} & \multirow{2}{*}{Gaze} & Call & 139 & 7870 & -5.28 & \textless{}.0001* \\
 &  &  & Non-call & 139 & 2570 &  &  \\
 & \multirow{2}{*}{$H_8$} & \multirow{2}{*}{Visit} & Call & 139 & 7939 & -5.28 & \textless{}.0001* \\
 &  &  & Non-call & 139 & 2501 &  &  \\
 \hline 
\multirow{4}{*}{\textbf{Experts}} & \multirow{2}{*}{$H_9$} & \multirow{2}{*}{Gaze} & Ctrl. Flow & 69 & 1246 & -1.13 & 0.25 \\
 &  &  & Non-Ctrl & 69 & 899 &  &  \\
 & \multirow{2}{*}{$H_{10}$} & \multirow{2}{*}{Visit} & Ctrl. Flow & 69 & 1487 & -2.71 & .007* \\
 &  &  & Non-Ctrl & 69 & 658 &  &  \\
 \hline 
\multirow{4}{*}{\textbf{Novices}} & \multirow{2}{*}{$H_{11}$} & \multirow{2}{*}{Gaze} & Ctrl. Flow & 139 & 5339 & -1.47 & 0.13 \\
 &  &  & Non-Ctrl & 139 & 3977 &  &  \\
 & \multirow{2}{*}{$H_{12}$} & \multirow{2}{*}{Visit} & Ctrl. Flow & 139 & 5961 & -2.83 & .005* \\
 &  &  & Non-Ctrl & 139 & 3355 &  & \\
 \hline
\end{tabular}%

}
\end{table}

\textit{Calls:}
We found evidence that developers read method calls more than other locations.  On average, the experts spent 44\% of their gaze time reading calls ($\approx$ 38\% of methods). Similarly, on average, novices spent 46\% of their gaze time reading calls ($\approx$ 39\% of methods). We define the four hypotheses ($H_5$, $H_6$, $H_7$, and $H_8$) regarding call terms as follows. 

\textit{$H_n$: For [experts / novices], the difference between the adjusted [gaze time / visit] metric for call terms and non-terms terms is not statistically  significant.}

We reject all the four hypotheses $H_5$ (medium effect size of $0.25$), $H_6$ (small effect size of $0.18$), $H_7$ (medium effect size of $0.31$), and $H_8$ (medium effect size of $0.31$) in Table \ref{tab:mainresult}.  Both experts and novices read call terms more heavily than non-call terms when the gaze time and number of visits are adjusted based on the number of terms.

\textit{Control Flow:}
We found no evidence that experts read control flow terms longer than other areas but they visit them more frequently.  On average, the experts spent 28\% of their gaze time reading control flow terms ($\approx$ 26\% of methods) while novices spent 26\% of their gaze time reading control flow terms  ($\approx$ 24\% of methods).  The following presents the definition of the four hypotheses ($H_9$, $H_{10}$, $H_{11}$, and $H_{12}$).

\textit{${H_n}$: For [experts / novices], the difference between the adjusted [gaze time / visit] metric for control-flow terms and control-flow terms is not statistically significant.}

We reject two hypotheses ($H_{10}$ with small effect size of $0.24$, and $H_{12}$ with small effect size of $0.18$) in Table \ref{tab:mainresult}.  This suggests that experts and novices revisit control flow terms but do not read them for a long period of time during each visit. 



\textit{Adjusted duration and visits based on number of characters:} 
Gaze time and number of visits for both call terms and control flow terms are adjusted based on the number of characters. This assists in determining if the significance  of results is due to  the complexity of the terms (number of characters) \cite{Rodeghero2015TSE,Liblit2006} or the importance of terms at that location.  For experts, when call terms are adjusted by number of characters, the difference between call terms and non-call terms is not statistically significant for the gaze time ($Z$ = -.828, $p$ = .4) and number of visits ($Z$ = -.332, $p$ = .47). On the other hand, for novices, when the gaze time and number of visits are adjusted based on the number of characters, the difference between call terms and non-call terms remains significant for both gaze time ($Z$= -4.46, $p$ \textless 0.001) and number of visits ($Z$ = -3.45, $p$ \textless 0.001). This  indicates that experts spend longer time on call terms due to their length (number of characters) while novices demand longer gaze time to comprehend call terms. Identifiers in all five systems were shown in camel case.

For both novices and experts, the statistical differences of number of visits remain significant when the results of control flow terms are adjusted based on the number of characters.  For experts, the statistical results for gaze time and number of visits are ($Z$= -1.82, $p$ = .068) and ($Z$= -3.38, $p$ = .001), respectively.  For novices, the statistical result for gaze time is ($Z$= -2.47, $p$ = 0.013) and for number of visits is ($Z$ = -3.66, $p$ \textless 0.001). Note that when gaze time of control flow terms is adjusted by number of characters, the difference between control flow terms and non-control flow terms become significant.  This indicates that, unlike experts, novices need  significant time to comprehend control flow terms.

Therefore, we conclude that calls and control flow terms are read differently. In particular, call  terms require longer gaze time and higher number of visits while control flow terms are visited more frequently by experts and novices. However, one cannot claim to have evidence to order these locations with respect to their importance (even though this is suggested in \cite{Rodeghero2014}). For example, experts spend a long time comprehending calls due to their complex identifiers. However, this long gaze time does not necessarily suggest that call terms are more important than other locations. If a method has ten call invocations, these calls may not be equally important. The same reasoning applies to control flow terms. A more detailed analysis of this is left as future work. 

\textbf{RQ1 Discussion on Similarities and Differences in Results with the Rodeghero study}: 
Rodeghero et al. observed that long identifiers (high number of characters) experienced long gaze time \cite{Rodeghero2015TSE} due to their complexity \cite{Liblit2006}. Similarly, we observe that experts read call terms significantly longer than non-call terms when adjusted by number of terms but the result becomes insignificant when time is adjusted by number of characters (no character adjustment was done in \cite{Rodeghero2015TSE}). This indicates that call terms required longer time due to the complexity of the terms (number of characters) \cite{Liblit2006}.


As seen in Table \ref{tab:collincomp}, these results do not reproduce results from the Rodeghero study \cite{Rodeghero2014}. They concluded that the signature is the most important location followed by call terms and next by control flow terms. However, in the study presented here, developers read and revisit call terms more heavily than other locations. Control flow terms are revisited more frequently than other terms. Finally, signature terms are revisited less frequently than other locations. Therefore, we conclude that developers read call terms the most followed by control flow terms and finally the signature.

\begin{table}[]
\caption{Comparing Results with the Rodeghero study{\cite{Rodeghero2014}}. Our results show developers read call terms the most followed by control flow and finally the signature. Higher rank is based on higher gaze time and number of visits.}

\label{tab:collincomp}
\resizebox{\columnwidth}{!}{%
\begin{tabular}{c c c}
\hline
\textbf{Rank} & \textbf{Rodeghero Ranking} & \textbf{This Study Ranking} \\ \hline
\textbf{1} & Signature & Call terms \\ 
\textbf{2} & Call terms & Control flow terms \\ 
\textbf{3} & Control flow terms & Signature \\ \hline
\end{tabular}%
}
\end{table}

We interpret the difference as follows. First, developers read methods differently when presented inside their inherent context versus in isolation. Developers tend to focus on the method's body more than its signature when the code is presented in a realistic environment \cite{Kevic2017}. In the Rodeghero study, when a developer formulates a hypothesis about a method and wants to step back, they have no choice but to look at the signature. However in our case, they can examine locations outside the method. Second, our study uses longer and more complex methods (the largest method is 80 LOC) compared to the Rodeghero study (the largest method 22 LOC). In fact, we found statistical evidence that the larger the method, the less likely experts fixate or revisit the signature. Furthermore, Rodeghero et al. concluded that control flow terms are read statistically less than other areas of a method \cite{Rodeghero2014}. Our result indicates that control flow terms in larger methods (rated to be harder to summarize by experts) are visited more frequently than control terms of smaller methods. 

\textbf{RQ1 Finding:} Experts and novices spend the longest time on reading call terms and visit call terms more than non-call terms. The next most read and visited locations are control flow terms and the method signature.

\subsection{RQ2 Results: Method Size }

We analyze eye-tracking data based on method size using the same approach in RQ1 by adjusting the gaze time and number of visits for each location (signature, body, calls, control flow). Then, methods are divided by their size - large (40-80 LOC), medium (23-39 LOC), and small (9-22 LOC).  For each  category, the eye movements on signature, call, and control flow statements are analyzed. These findings are reported for experts and novices. The hypotheses are:

\textit{$H_n$: With respect to [experts / novices] for [large / medium / small] methods, the difference between the adjusted [gaze time/visit] metric for [“signature and the body” / “calls and other areas” / “control flow and other areas”] is not statistically significant.}

\textit{Experts:}We only report the statistical significant results due to space limitations. Experts revisit the body of a method statistically more frequently than the signature in case of large methods ($n = 18$, $Z = -2.94$, $p = .003$ with a large effect size of $0.49$).  No statistical differences were found in case of medium ($n = 22$, $Z$ = -2.22, $p = .026$) and small ($n = 29$, $Z$ = -1.22, $p = .22$) methods. Therefore, we conclude that as the size of a method increases, experts visit method bodies more often than method signatures.
For control flow and call terms, no statistical differences were found in all cases (small, medium and large methods).

\textit{Novices:} Novices read the signature significantly less than the body in case of small methods  - in terms of gaze time ($n = 68$, $Z = -3.15$, $p = .002$ with a medium effect size of $0.27$) and number of visits ($Z = -3.41$, $p = .001$ with a medium effect size of $0.29$).
Novices spent a significant amount of time reading call terms except for large methods. They read call terms significantly more than non-call terms in case of small methods - in terms of gaze time ($n = 68$, $Z = -3.89$, $p$ = \textless$.0001$ with a medium effect size of $0.33$) and number of visits ($Z = -3.94$, $p = .0001$ with a medium effect size of $0.37$).
Novices read call terms significantly more than non-call terms in case of medium methods in terms of gaze time ($n = 29$, $Z = -2.69$, $p = .007$ with a medium effect size of $0.35$) and number of visits ($Z = -2.86$, $p = .002$ with a medium effect size of $0.37$).
After manually checking novices' answers we observed that they wrote their summaries from the signature and avoided reading the code especially for large methods. 
For control flow terms, no statistical differences were found in small, medium and large methods.


\textbf{RQ2 Finding:} We found statistical evidence that experts tend to revisit the body of the method more frequently than its signature when the size of the method increases.


\subsection{RQ3 Results: Reading Outside of the Summarized Method }

As mentioned earlier, since we used the Eclipse plugin, iTrace \cite{Shaffer2015}, the participants were able to context switch to any file if they needed to.  They worked in a very realistic setting where all the code was provided to them. When a developer starts summarizing a method and then switches to some location outside of the scope of the assigned method, this may be an indication that they need more information to understand the method or to confirm an assumption.  Accordingly, they spend some time looking for clues that might be found somewhere else, either within the same file or outside the file that has the assigned method to be summarized.  These possible scenarios are examined for experts and novices.

Both experts and novices read outside of the assigned method in 70\% (213 out of 257) of the tasks. The average time spent by experts to examine locations outside of the method is generally lower than the average time spent by novices. On average, novices spent about 6\% of the time examining outside locations whereas experts spent about 4\%. We further inspected the summary responses collected from the participants after the experiment and compared their responses to the average time spent outside the method.  All methods for which participants spent more than average time outside the method, at least one of the participants described the method (or one of its components such as a data member or method call) as vague, used the word ``unsure" in their summary, or rated the method as difficult in a post-questionnaire on task difficulty.

\textbf{RQ3 Discussion}: We discuss three out-of-scope locations that a participant may have examined while summarizing an assigned method. These locations are: class declarations, data members, and close approximation (i.e., statements that belong to methods that precede or follow the assigned method).


For both experts and novices, the largest duration spent outside the method scope is to read in close approximation.  
Specifically, we observed that when a participant is reading the first few lines of the assigned method, he or she also looks at the methods or lines that precede the method (with respect to the location in the file).  Similarly, when a participant is inspecting the last few statements of an assigned method, he or she may also look at those lines that appeared after the method in the file.  This observation confirms the result of \cite{Kevic2015}.  However, it does not provide insight that the participants might have got help from looking at those nearby locations.

Data members and class declarations are the next common locations that experts and novices examine outside the assigned method.  
After manually examining the data, we noticed that locating data members and class declarations is mostly done by scrolling up or down rather than using the search option.  This might be the case as the class declaration is always found at the beginning of the file and data members are found either at the beginning or the end of the source file. Also, scrolling up or down to look for information is used usually when the file that contains the assigned method is short. Using the search option is noticed in cases of large files. 
On average, experts spent 11\% of the gaze time reading data members and 4\% of gaze time reading the class declaration. Novices spent 1\% of the gaze time reading data members and 1\% of gaze time reading class declaration. 
Furthermore, we observed one expert and one novice switch from a statement with a method call (within the assigned method to summarize) to an outside location that follows the sequence of the methods captured in (or represented by) the call graph.

We posit that for both experts and novices, reading outside the method scope is more related to developer preference than their level of expertise. We found three experts and four novices who read outside a method more often than others. 
Finally, a random scan of other methods inside the file is a behavior that is carried more frequently by novices while experts tend to focus on the assigned methods and read outside the method if they are looking for information such as the class or data member declarations.

\textbf{RQ3 Finding:} All participants looked outside the scope of the assigned methods at least once during the experiment.  On average, they spend $94.23\%$ of their time examining locations within the assigned method and $5.77\%$ of the time is spent examining other locations outside the scope of the method. All five experts examine data members at least once and three experts read the class declarations. Searching for them by scrolling is more common than using the search option.

\subsection{RQ4 Results: Source Lines Read / Used in Summarization}


For this RQ, we choose to focus on tasks completed by experts only because we observe that experts write concise and high-quality summaries compared to novices. Although the analysis of novices’ data can help understand students’ reading behavior and eventually can contribute in better teaching students, such analysis is out of the scope of this paper. 
Unlike RQ1 and RQ2 that groups all terms for each statement type (i.e., control flow), we study each line as whole in RQ3. Therefore, for each source code line read by an expert, the total gaze time is computed. The gaze time of a line is the total amount of time spent reading any term on that line. This analysis is done on 69 samples (spanning a total of 1168 lines read by all experts). 

\begin{figure}
\centering
\includegraphics[scale=0.50]{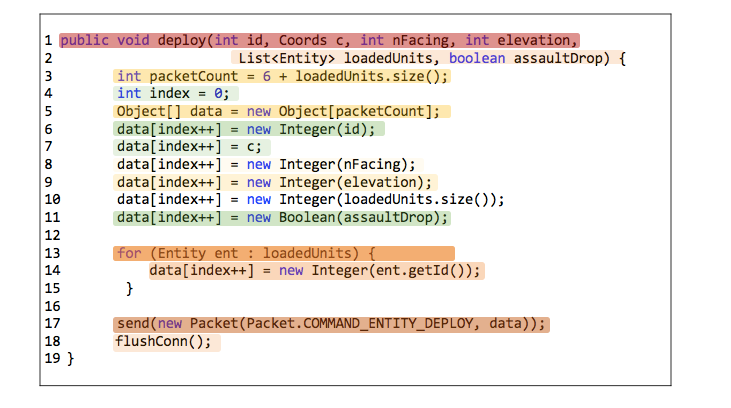}
\caption{Heat map generated for one method and one participant. Longest gaze time (red lines), long gaze time (orange lines), medium gaze time (yellow lines), short gaze time (light green lines), shortest gaze time but not zero (green lines), and zero gaze time (uncolored lines).}
\label{fig:heatmap}
\end{figure}

We observed that $15\%$ of lines with no gaze data are lines with no identifiers (e.g., a open/close brace, return false; or having the `else' keyword) as no higher level information can be obtained from them. However, a high number of identifiers on a line may not lead directly to long gaze time. See Figure \ref{fig:heatmap} for a heat map (a common graphical representation of eye-tracking data) of one sample. 
Although line 18 is the shortest line in the method, this participant spent about $3,832 ms$  on this line which is 6\% of the total gaze time. On the other hand, longer lines could have shorter gaze times such as line 6 ($933 ms$) and line 11 ($1,832 ms$).


\begin{figure*}[tp]
\centering
\includegraphics[scale=0.25]{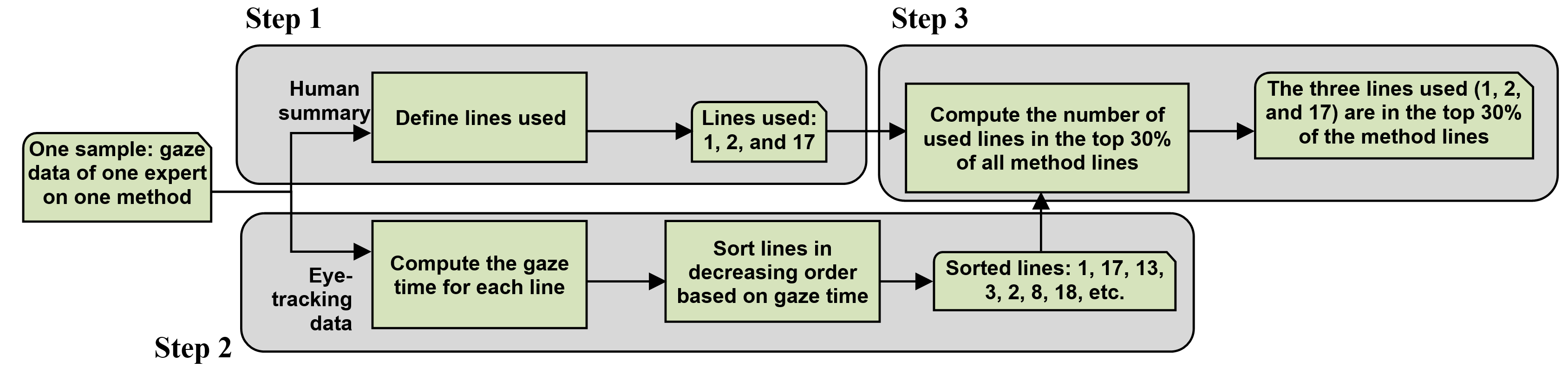}
\caption{Illustrating the three steps performed to answer RQ 4 for the method in Figure \ref{fig:heatmap}.}
\label{fig:steps}
\end{figure*}

We first study the impact of line length on the gaze time to determine if we need to adjust the gaze time based on line length to make a fair comparison for analysis. 
\textit{First}, for each data sample, according to the gaze time, lines are sorted in decreasing order (from largest to smallest). Therefore, the line with the largest gaze time has a rank of one and the line with the lowest gaze time has a rank equal to $n$, where $n$ is the number of lines in a data sample. \textit{Second}, for the same data sample, we adjust the gaze time for each source code line based on the number of identifiers. For example, if a participant spent $100 ms$ reading a line and the line has 4 identifiers, then the adjusted duration is $100/4 = 25 ms$. \textit{Third}, each line is assigned another rank based on the adjusted gaze time as we did with the actual time in the first step.
Finally, the original list is compared to the adjusted duration using Kendall's tau coefficient ($\uptau$), that evaluates the degree of concordance between two sets of ranked data \cite{siegel1956nonparametric,Abdi2007} - a suitable test in this case. If Kendall's tau is 1, then the two lists are identical, -1 indicates that one is the reverse of the other. 
Since critical value of Kendall's tau depends on n (number of lines in each sample), samples that has less than five lines are eliminated because n is not large enough \cite{Abdi2007}.
We found significant agreement between the line ranks of the actual gaze time and adjusted gaze time in 90\% (61 out of 68) of the samples. Additionally, we found  insignificant agreement between the line ranks of the actual gaze time and adjusted gaze time in 10\% (7 out of 68) of the samples.
This strongly suggests that the length of a line does not significantly impact the amount of time spent reading the line. Consequently, for each line, we can use either the actual gaze time or the adjusted gaze time. We use actual gaze time for the rest of the analysis.



We analyze expert summaries in order to understand what developers seek in a method summary. These observations can ultimately produce recommendations and guidelines regarding source code summarization and source code comprehension.
Lines that experts used to write their summaries are first manually identified. Then, the gaze time of these lines are analyzed. Figure \ref{fig:steps} demonstrates the three steps performed for the method shown in Figure \ref{fig:heatmap}.   
We first determine lines used to write the summary for a method. This is done by manually examining the expert summaries. Consider the method presented in Figure \ref{fig:heatmap}. The summary written by \textit{expert A} is ``Deploys units for the game, communicating this over a network connection". The two words `Deploys' and `units' appear together in the signature on line 1 and 2. The main verb `deploy' also appears on line 17. Therefore, the list of lines used are: lines 1, 2 and 17. The second part of the summary ``communicating this over a network connection" suggests the usage of line 17 although none of the words are directly used from the method call in line 17. 
The summary of the same method written by \textit{expert B} is ``Transmit game unit object data positioning over the network and flush the connection" which suggests the use of lines 5, 17 and 18. Line 5 is added because this participant specifies that type and the name of the data being sent ``object data". Finally, both experts use the word ``game" although it does not appear in the method. They might have gained this information by reading the description of the system. Finally, if a phrase used by the developer is found on multiple lines (e.g., a call that appears three times in the method), the three lines are included in the lines-used list as long as the gaze duration is greater than zero. Of the $1,168$ lines, 181 lines are used by 5 experts. On average, experts use content from three lines to write their summaries for a method. 

After these lines are determined, they are sorted in decreasing order. 
Finally, a comparison is made to determine if lines used (lines 1, 2 and 17 in our example) are lines that the expert read. As shown in Figure \ref{fig:steps}, of the 20 lines, line 1 and 17 are the most read lines and line 2 is the fifth most read line. In this sample, for a used line to be in the top 30\% read lines, the line should be in the (20 lines)/ 3 =  $6^{th}$ position or higher.  Therefore, \textit{expert A} wrote her summary using lines that she read. These three steps are repeated for all 69 experts’ samples for the 181 lines used. 
We found that 70\% (127 out of 181) of used lines are located in the top 30\% of lines with the highest gaze time. 87\% (157 out of 181 ) of used lines are located in the top 50\% of lines with the highest gaze time and finally 100\% of used lines are located in the top 86\% of lines with the highest gaze time.





A line from a method can appear more than once in the set of 181 lines used if more than one expert used it. To examine if participants use similar lines, we analyze methods that are summarized by at least two experts. 15 methods are summarized by 3-5 experts who use 85 unique lines to write their summaries. 
54\% of lines are used by one expert only and 46\% of lines used by at least two experts. We interpret this as experts tend to use similar lines but the level of details varies.



\textbf{RQ4 Discussion}:
In the list of 181 used lines, there are 124 unique lines from 15 methods. One of the authors manually examined all 124 unique lines used by experts to determine the type of these lines. We found that 119 (out of 124) of lines used by experts belong to one of the summary units proposed in previous literature on source code summarization \cite{Sridhara2010,McBurney2014,Abid2015,Sridhara2011a,Rodeghero2014}. \textit{Summary units} are sets of source code lines that are used to generate an automatic summary of the method. These proposed tools select the lines from the method based on linguistic and/or static analysis.
Table \ref{tab:mapping} demonstrates summary unit categories along with summarization lines introduced in the literature and how many of the lines used by experts (in our study) fall into these categories. 
For example, the method name is commonly used to summarize a method \cite{McBurney2014,Abid2015,Rodeghero2014} as it describes the goal of a method. 
In the example shown in Fig. \ref{fig:heatmap}, the \textit{last void call} summarization unit is used (the main summary units are lines 17 and 18). Additionally, one of the \textit{secondary summary units} can be used to support one of the main summary units. For example, line 5 - variable declaration - is used to add more information to line 17. We conclude that our findings validate the literature at a fine-grained level. The lines used by experts are actually the same lines that past literature has proposed. \cite{Sridhara2010,McBurney2014,Abid2015,Sridhara2011a,Rodeghero2014}.

\begin{table*}
\centering
\caption{The type of source code lines used by experts to write their summaries. These lines match source code lines proposed by the literature in automatic summarization/documentation}
\label{tab:mapping}
\resizebox{\textwidth}{!}{%
\begin{tabular}{rclcc}
\hline
\multicolumn{2}{c}{\textbf{Category of Summary Unit}} & \multicolumn{1}{c}{\textbf{Statements Type}} & \textbf{Statement Subtype} & \textbf{\# occurrences} \\ \hline
\multicolumn{2}{c|}{\textbf{Method Signature}} & \multicolumn{1}{l|}{(1) Method Name and parameters. McBurney'14 \cite{McBurney2014}, Rodeghero'14 \cite{Rodeghero2014}, Abid'15 \cite{Abid2015}} & \multicolumn{1}{c|}{} & 13 \\ \hline
\multicolumn{1}{r|}{\multirow{16}{*}{\begin{sideways}\textbf{Method Body}\end{sideways}}} & \multicolumn{1}{c|}{\multirow{7}{*}{\textbf{Main Summary Units}}} & \multicolumn{1}{l|}{(2) Last void call. Sridhara'10 \cite{Sridhara2010}} & \multicolumn{1}{c|}{} & 15 \\ \cline{3-5} 
\multicolumn{1}{r|}{} & \multicolumn{1}{c|}{} & \multicolumn{1}{l|}{(3) Same action (call name inside method body same as method name) Sridhara'10 \cite{Sridhara2010}} & \multicolumn{1}{c|}{} & 13 \\ \cline{3-5} 
\multicolumn{1}{r|}{} & \multicolumn{1}{c|}{} & \multicolumn{1}{l|}{\multirow{2}{*}{\begin{tabular}[c]{@{}l@{}}(4) Modifying a data member by {Abid'15 \cite{Abid2015}}\end{tabular}}} & \multicolumn{1}{c|}{call} & 9 \\ \cline{4-5} 
\multicolumn{1}{r|}{} & \multicolumn{1}{c|}{} & \multicolumn{1}{l|}{} & \multicolumn{1}{c|}{assignment} & 2 \\ \cline{3-5} 
\multicolumn{1}{r|}{} & \multicolumn{1}{c|}{} & \multicolumn{1}{l|}{\multirow{2}{*}{\begin{tabular}[c]{@{}l@{}}(5) Modifying a parameter by {Sridhara'11 \cite{Sridhara2011b}, Abid'15 \cite{Abid2015}}\end{tabular}}} & \multicolumn{1}{c|}{call} & 4 \\ \cline{4-5} 
\multicolumn{1}{r|}{} & \multicolumn{1}{c|}{} & \multicolumn{1}{l|}{} & \multicolumn{1}{c|}{assignment} & 0 \\ \cline{3-5} 
\multicolumn{1}{r|}{} & \multicolumn{1}{c|}{} & \multicolumn{1}{l|}{(6) Return. Sridhara'10 \cite{Sridhara2010}, Abid'15 \cite{Abid2015}} & \multicolumn{1}{c|}{} & 2 \\ \cline{2-5} 
\multicolumn{1}{r|}{} & \multicolumn{1}{c|}{\multirow{6}{*}{\textbf{Secondary summary units}}} & \multicolumn{1}{l|}{\multirow{2}{*}{\begin{tabular}[c]{@{}l@{}}(7) Control flow that controls one of the main summary units {Sridhara'10 \cite{Sridhara2010}, Abid'15 \cite{Abid2015}}\end{tabular}}} & \multicolumn{1}{c|}{If/switch} & 19 \\ \cline{4-5} 
\multicolumn{1}{r|}{} & \multicolumn{1}{c|}{} & \multicolumn{1}{l|}{} & \multicolumn{1}{c|}{Loops} & 2 \\ \cline{3-5} 
\multicolumn{1}{r|}{} & \multicolumn{1}{c|}{} & \multicolumn{1}{l|}{\multirow{2}{*}{\begin{tabular}[c]{@{}l@{}}(8) Modifying the Returned value by {Sridhara'10 \cite{Sridhara2010}, Abid'15 \cite{Abid2015}}\end{tabular}}} & \multicolumn{1}{c|}{call} & 6 \\ \cline{4-5} 
\multicolumn{1}{r|}{} & \multicolumn{1}{c|}{} & \multicolumn{1}{l|}{} & \multicolumn{1}{c|}{assignment} & 12 \\ \cline{3-5} 
\multicolumn{1}{r|}{} & \multicolumn{1}{c|}{} & \multicolumn{1}{l|}{\multirow{2}{*}{\begin{tabular}[c]{@{}l@{}}(9) Variable declaration and/or initialization that used in one of the main summary unit {Sridhara'10 \cite{Sridhara2010}}\end{tabular}}} & \multicolumn{1}{c|}{Call} & 16 \\ \cline{4-5} 
\multicolumn{1}{r|}{} & \multicolumn{1}{c|}{} & \multicolumn{1}{l|}{} & \multicolumn{1}{c|}{assignment} & 2 \\ \cline{2-5} 
\multicolumn{1}{r|}{} & \multicolumn{1}{c|}{\textbf{Additional Information}} & \multicolumn{1}{l|}{(10) Variable declaration and/or initialization used in one of the secondary summary units. Abid'15 \cite{Abid2015}} & \multicolumn{1}{c|}{if-stmt} & 4 \\ \cline{2-5} 
\multicolumn{1}{r|}{} & \multicolumn{1}{c|}{\multirow{2}{*}{\textbf{Others}}} & \multicolumn{1}{l|}{(11) Void-call (calls that don't return a value)} & \multicolumn{1}{c|}{} & 3 \\ \cline{3-5} 
\multicolumn{1}{r|}{} & \multicolumn{1}{c|}{} & \multicolumn{1}{l|}{(12) If-statement (predicate)} & \multicolumn{1}{c|}{} & 2 \\ \hline
\multicolumn{4}{c}{\textbf{Total}} & 124 \\ \hline
\end{tabular}%
}
\end{table*}

\textbf{RQ4 Finding:} We found strong evidence that the gaze time of a line can predict if a line should be in a method summary. Our analysis reveals that developers tend to use source code lines that they read the most (highest gaze time) when they write their own summaries. We found that 46\% of lines used are found in summaries written by at least two experts.  

\section{Threats to Validity}
\label{sec:threats}
\textit{Internal Validity}: When a developer writes a summary for a method from a class, he/she may build some knowledge about the class.  This might affect the time and the effort to understand other methods from the same class. To mitigate this, each developer was asked to summarize methods from different unrelated classes. To avoid fatigue, we kept the number of methods to be summarized to 15 with the goal of having the study completed in about an hour. To reduce the overhead in browsing many systems, we limit the number of systems for each developer to three. As developers mostly rely on comments to understand methods \cite{Crosby1990}, we remove all comments from the source code (similar to the Rodeghero study). This was necessary as the goal of the study is to examine source code statements that developers focus on when they write their \textit{own} summaries. 
\textit{Construct Validity}: Developers have different IDE environment preferences that might affect their performance. We kept the default barebones Eclipse syntax highlighting preferences for all participants. In addition, code folding was not allowed to reduce any confounding effects. None of the participants complained about this setting or the highlighting used. 
Lines identified in RQ4 are determined by one of the authors. Future work will involve more than one author and will report agreement. 
\textit{External validity} refers to generalizing the results to the target population.  Our experts were comparable to industry programmers. 
To support \textit{conclusion validity}, we use appropriate statistical tests to match our data assumptions, i.e., using Wilcoxon signed-rank for paired non-parametric data and also report effect sizes. 


\section{Implications}
\label{sec:discussion}

These results can be directly applied to automatic summarization approaches. 13 out of 26 (See Table \ref{tab:mapping}) of the signatures are used by at least one expert to write their summaries. We observe that the expert starts the summary with the verb from the signature to explain the main action of the method. Then, adds more details from a set of selected calls and other lines from the body. We suggest the following guidance to summarization tool authors.
\begin{itemize}
\item Use the signature to reflect the main action of a method. 
\item Select one or more main summary units (see Table \ref{tab:mapping}). 
\textit{Last void call} is the main summary unit that is often used (15 times) by experts as methods often perform a set of steps to accomplish a final action \cite{Sridhara2010}. Therefore, we advise to include the \textit{last void call} of a method to the method summary. The \textit{same action} \cite{Sridhara2010} summary unit should also be used when applicable. 
Finally, 
one can distinguish between three types of summary units: (4) modifying a data member; (5) modifying a parameter; and (6) returning a computed value. 

\item Select one or more secondary summary units related to the selected main summary unit as shown in Table \ref{tab:mapping}. 
\end{itemize}

The results of this study can benefit other kinds of summarization such as automatic code folding. During maintenance tasks \cite{Kevic2017}, not all lines are equally important. For example, developers tend to ignore catch blocks as they are typically not important for summarization.  Furthermore, experts tend to focus on about 14\% of a method to write a summary. This finding can be used by folding those lines that are commonly ignored by experts.  Additionally, during a maintenance task, the automatic documentation summary of a method can be displayed, then using eye-tracking data, one can examine if developers tend to read and/or use the provided documentation.

\section{Conclusions and Future Work}
\label{conclusion}

This paper presents an eye-tracking study of 18 experts and novices reading and summarizing Java methods.  The study uses methods of different sizes to investigate their influence on developers' reading behavior. 
The results reveal that the signature of a method is statistically less visited by experts and novices. They also spend a significant amount of gaze time and have higher gaze visits when they read call terms. On the other hand, both experts and novices tend to revisit control flow terms rather than read them for a long period. Furthermore, analyzing lines that experts read during summarization can uncover important lines for automatic summarization.

The main take home message of the results are that conducting eye-tracking studies on a single screen of code does not always generalize to how a developer behaves in a real work environment.  This has been a limitation of eye-tracking software but with improvements in infrastructure, researchers can now study developers in a more realistic setting resulting in better understanding of how they actually solve software engineering tasks. As future work, we plan to replicate other eye tracking studies to see if the results generalize when conducted outside a constrained single-screen study environment.


\section*{Acknowledgment}
We are grateful to all the participants who took part in this study. This work is supported in part by grants from the National Science Foundation under grant numbers CCF 18-55756 and CCF 15-53573.


%
\balance 

\bibliographystyle{IEEEtran}
\setlength{\parindent}{0pt}
\bibliography{refs}

\begin{thebibliography}{10}
\providecommand{\url}[1]{#1}
\csname url@samestyle\endcsname
\providecommand{\newblock}{\relax}
\providecommand{\bibinfo}[2]{#2}
\providecommand{\BIBentrySTDinterwordspacing}{\spaceskip=0pt\relax}
\providecommand{\BIBentryALTinterwordstretchfactor}{4}
\providecommand{\BIBentryALTinterwordspacing}{\spaceskip=\fontdimen2\font plus
\BIBentryALTinterwordstretchfactor\fontdimen3\font minus
  \fontdimen4\font\relax}
\providecommand{\BIBforeignlanguage}[2]{{%
\expandafter\ifx\csname l@#1\endcsname\relax
\typeout{** WARNING: IEEEtran.bst: No hyphenation pattern has been}%
\typeout{** loaded for the language `#1'. Using the pattern for}%
\typeout{** the default language instead.}%
\else
\language=\csname l@#1\endcsname
\fi
#2}}
\providecommand{\BIBdecl}{\relax}
\BIBdecl

\bibitem{ko2006}
A.~J. Ko, B.~A. Myers, M.~J. Coblenz, and H.~H. Aung, ``An exploratory study of
  how developers seek, relate, and collect relevant information during software
  maintenance tasks,'' \emph{IEEE Transactions on Software Engineering},
  vol.~32, no.~12, pp. 971--987, 2006.

\bibitem{LaToza2006}
T.~D. LaToza, G.~Venolia, and R.~DeLine, ``Maintaining mental models: A study
  of developer work habits,'' in \emph{Proceedings of the 28th International
  Conference on Software Engineering}, ser. ICSE '06, 2006, pp. 492--501.

\bibitem{Fluri2007}
B.~Fluri, M.~Wursch, and H.~C. Gall, ``Do code and comments co-evolve? on the
  relation between source code and comment changes,'' in \emph{14th Working
  Conference on Reverse Engineering (WCRE 2007)}, 2007, pp. 70--79.

\bibitem{Crosby1990}
M.~E. Crosby and J.~Stelovsky, ``How do we read algorithms? a case study,''
  \emph{Computer}, vol.~23, no.~1, pp. 25--35, 1990.

\bibitem{deSouza2005}
S.~C.~B. de~Souza, N.~Anquetil, and K.~M. de~Oliveira, ``A study of the
  documentation essential to software maintenance,'' in \emph{Proceedings of
  the 23rd Annual International Conference on Design of Communication:
  Documenting \&Amp; Designing for Pervasive Information}, ser. SIGDOC '05,
  2005, pp. 68--75.

\bibitem{Sridhara2010}
G.~Sridhara, E.~Hill, D.~Muppaneni, L.~Pollock, and K.~Vijay-Shanker, ``Towards
  automatically generating summary comments for java methods,'' in
  \emph{Proceedings of the IEEE/ACM International Conference on Automated
  Software Engineering}, ser. ASE '10, 2010, pp. 43--52.

\bibitem{McBurney2014}
P.~W. McBurney and C.~McMillan, ``Automatic documentation generation via source
  code summarization of method context,'' in \emph{Proceedings of the 22Nd
  International Conference on Program Comprehension}, ser. ICPC 2014, 2014, pp.
  279--290.

\bibitem{Haiduc2010}
S.~Haiduc, J.~Aponte, L.~Moreno, and A.~Marcus, ``On the use of automated text
  summarization techniques for summarizing source code,'' in \emph{2010 17th
  Working Conference on Reverse Engineering}, 2010, pp. 35--44.

\bibitem{Eddy2013}
B.~P. Eddy, J.~A. Robinson, N.~A. Kraft, and J.~C. Carver, ``Evaluating source
  code summarization techniques: Replication and expansion,'' in \emph{2013
  21st International Conference on Program Comprehension (ICPC)}, 2013, pp.
  13--22.

\bibitem{Moreno2013}
L.~Moreno, J.~Aponte, G.~Sridhara, A.~Marcus, L.~Pollock, and K.~Vijay-Shanker,
  ``Automatic generation of natural language summaries for java classes,'' in
  \emph{2013 21st International Conference on Program Comprehension (ICPC)},
  2013, pp. 23--32.

\bibitem{Abid2015}
N.~J. Abid, N.~Dragan, M.~L. Collard, and J.~I. Maletic, ``Using stereotypes in
  the automatic generation of natural language summaries for c++ methods,'' in
  \emph{2015 IEEE International Conference on Software Maintenance and
  Evolution (ICSME)}, 2015, pp. 561--565.

\bibitem{Sridhara2011a}
G.~Sridhara, L.~Pollock, and K.~Vijay-Shanker, ``Automatically detecting and
  describing high level actions within methods,'' in \emph{Proceedings of the
  33rd International Conference on Software Engineering}, ser. ICSE '11, 2011,
  pp. 101--110.

\bibitem{Sridhara2011b}
------, ``Generating parameter comments and integrating with method
  summaries,'' in \emph{2011 IEEE 19th International Conference on Program
  Comprehension}, 2011, pp. 71--80.

\bibitem{Badihi2017}
S.~Badihi and A.~Heydarnoori, ``Crowdsummarizer: Automated generation of code
  summaries for java programs through crowdsourcing,'' \emph{IEEE Software},
  vol.~34, no.~2, pp. 71--80, 2017.

\bibitem{wan2018improving}
Y.~Wan, Z.~Zhao, M.~Yang, G.~Xu, H.~Ying, J.~Wu, and S.~Y. Philip, ``Improving
  automatic source code summarization via deep reinforcement learning,'' in
  \emph{The 33rd IEEE/ACM International Conference on Automated Software
  Engineering}, 2018.

\bibitem{Moreno2017}
L.~Moreno and A.~Marcus, ``Automatic software summarization: the state of the
  art,'' in \emph{2017 IEEE/ACM 39th International Conference on Software
  Engineering Companion (ICSE-C)}, 2017, pp. 511--512.

\bibitem{Rodeghero2014}
P.~Rodeghero, C.~McMillan, P.~W. McBurney, N.~Bosch, and S.~D'Mello,
  ``Improving automated source code summarization via an eye-tracking study of
  programmers,'' in \emph{Proceedings of the 36th International Conference on
  Software Engineering}, ser. ICSE 2014, 2014, pp. 390--401.

\bibitem{Shaffer2015}
T.~R. Shaffer, J.~L. Wise, B.~M. Walters, S.~C. M\"{u}ller, M.~Falcone, and
  B.~Sharif, ``itrace: Enabling eye tracking on software artifacts within the
  ide to support software engineering tasks,'' in \emph{Proceedings of the 2015
  10th Joint Meeting on Foundations of Software Engineering}, ser. ESEC/FSE
  2015, 2015, pp. 954--957.

\bibitem{Guarnera18iTrace}
D.~T. Guarnera, C.~A. Bryant, A.~Mishra, J.~I. Maletic, and B.~Sharif,
  ``itrace: Eye tracking infrastructure for development environments,'' in
  \emph{Proceedings of the 2018 ACM Symposium on Eye Tracking Research \&
  Applications}, ser. ETRA '18, 2018, pp. 105:1--105:3.

\bibitem{Sharafi2015}
Z.~Sharafi, Z.~Soh, and Y.~Gu\'eh\'eneuc, ``A systematic literature review on
  the usage of eye-tracking in software engineering,'' \emph{Information and
  Software Technology}, vol.~67, pp. 79 -- 107, 2015.

\bibitem{Sharif2017}
B.~Sharif, J.~Meinken, T.~Shaffer, and H.~Kagdi, ``Eye movements in software
  traceability link recovery,'' \emph{Empirical Software Engineering}, vol.~22,
  no.~3, pp. 1063--1102, Jun 2017.

\bibitem{Sharafi2015d}
Z.~Sharafi, T.~Shaffer, B.~Sharif, and Y.~Gu\'eh\'eneuc, ``Eye-tracking metrics
  in software engineering,'' in \emph{2015 Asia-Pacific Software Engineering
  Conference (APSEC)}, 2015, pp. 96--103.

\bibitem{Obaidellah2018}
U.~Obaidellah, M.~Al~Haek, and P.~C.-H. Cheng, ``A survey on the usage of
  eye-tracking in computer programming,'' \emph{ACM Comput. Surv.}, vol.~51,
  no.~1, pp. 5:1--5:58, Jan. 2018.

\bibitem{Barik2017}
T.~Barik, J.~Smith, K.~Lubick, E.~Holmes, J.~Feng, E.~Murphy-Hill, and
  C.~Parnin, ``Do developers read compiler error messages?'' in
  \emph{Proceedings of the 39th International Conference on Software
  Engineering}, ser. ICSE '17, 2017, pp. 575--585.

\bibitem{Bednarik2006}
R.~Bednarik and M.~Tukiainen, ``An eye-tracking methodology for characterizing
  program comprehension processes,'' in \emph{Proceedings of the 2006 Symposium
  on Eye Tracking Research \&Amp; Applications}, ser. ETRA '06, 2006, pp.
  125--132.

\bibitem{Bednarik2008}
------, ``Temporal eye-tracking data: Evolution of debugging strategies with
  multiple representations,'' in \emph{Proceedings of the 2008 Symposium on Eye
  Tracking Research \&\#38; Applications}, ser. ETRA '08, 2008, pp. 99--102.

\bibitem{Busjahn2015}
T.~Busjahn, R.~Bednarik, A.~Begel, M.~Crosby, J.~H. Paterson, C.~Schulte,
  B.~Sharif, and S.~Tamm, ``Eye movements in code reading: Relaxing the linear
  order,'' in \emph{2015 IEEE 23rd International Conference on Program
  Comprehension}, 2015, pp. 255--265.

\bibitem{Rodeghero2015ESEM}
P.~Rodeghero and C.~McMillan, ``An empirical study on the patterns of eye
  movement during summarization tasks,'' in \emph{2015 ACM/IEEE International
  Symposium on Empirical Software Engineering and Measurement (ESEM)}, vol.~00,
  2015, pp. 1--10.

\bibitem{Jbara2017}
A.~Jbara and D.~G. Feitelson, ``How programmers read regular code: a controlled
  experiment using eye tracking,'' \emph{Empirical Software Engineering},
  vol.~22, no.~3, pp. 1440--1477, Jun 2017.

\bibitem{Lee2017}
S.~Lee, D.~Hooshyar, H.~Ji, K.~Nam, and H.~Lim, ``Mining biometric data to
  predict programmer expertise and task difficulty,'' \emph{Cluster Computing},
  Jan 2017.

\bibitem{Uwano2006}
H.~Uwano, M.~Nakamura, A.~Monden, and K.-i. Matsumoto, ``Analyzing individual
  performance of source code review using reviewers' eye movement,'' in
  \emph{Proceedings of the 2006 Symposium on Eye Tracking Research \&Amp;
  Applications}, ser. ETRA '06, 2006, pp. 133--140.

\bibitem{Sharif2012}
B.~Sharif, M.~Falcone, and J.~I. Maletic, ``An eye-tracking study on the role
  of scan time in finding source code defects,'' in \emph{Proceedings of the
  Symposium on Eye Tracking Research and Applications}, ser. ETRA '12, 2012,
  pp. 381--384.

\bibitem{Begel2018}
A.~Begel and H.~Vrzakova, ``Eye movements in code review,'' in
  \emph{Proceedings of the Workshop on Eye Movements in Programming}, ser. EMIP
  '18, 2018, pp. 5:1--5:5.

\bibitem{Rodeghero2015TSE}
P.~Rodeghero, C.~Liu, P.~W. McBurney, and C.~McMillan, ``An eye-tracking study
  of java programmers and application to source code summarization,''
  \emph{IEEE Transactions on Software Engineering}, vol.~41, no.~11, pp.
  1038--1054, 2015.

\bibitem{Kevic2015}
K.~Kevic, B.~M. Walters, T.~R. Shaffer, B.~Sharif, D.~C. Shepherd, and
  T.~Fritz, ``Tracing software developers' eyes and interactions for change
  tasks,'' in \emph{Proceedings of the 2015 10th Joint Meeting on Foundations
  of Software Engineering}, ser. ESEC/FSE 2015, 2015, pp. 202--213.

\bibitem{Kevic2017}
K.~Kevic, B.~Walters, T.~Shaffer, B.~Sharif, D.~C. Shepherd, and T.~Fritz,
  ``Eye gaze and interaction contexts for change tasks - observations and
  potential,'' \emph{Journal of Systems and Software}, vol. 128, pp. 252--266,
  2017.

\bibitem{Wang2015}
X.~Wang, L.~Pollock, and K.~Vijay-Shanker, ``Developing a model of loop actions
  by mining loop characteristics from a large code corpus,'' in \emph{2015 IEEE
  International Conference on Software Maintenance and Evolution (ICSME)},
  2015, pp. 51--60.

\bibitem{Wang2017}
------, ``Automatically generating natural language descriptions for
  object-related statement sequences,'' in \emph{2017 IEEE 24th International
  Conference on Software Analysis, Evolution and Reengineering (SANER)}, 2017,
  pp. 205--216.

\bibitem{Dragan2006}
N.~Dragan, M.~L. Collard, and J.~I. Maletic, ``Reverse engineering method
  stereotypes,'' in \emph{2006 22nd IEEE International Conference on Software
  Maintenance}, 2006, pp. 24--34.

\bibitem{Dragan2010}
------, ``Automatic identification of class stereotypes,'' in \emph{2010 IEEE
  International Conference on Software Maintenance}, 2010, pp. 1--10.

\bibitem{Abid2017}
N.~Abid, N.~Dragan, M.~L. Collard, and J.~I. Maletic, ``The evaluation of an
  approach for automatic generated documentation,'' in \emph{2017 IEEE
  International Conference on Software Maintenance and Evolution (ICSME)},
  2017, pp. 307--317.

\bibitem{Mcburney2016}
P.~W. Mcburney and C.~Mcmillan, ``An empirical study of the textual similarity
  between source code and source code summaries,'' \emph{Empirical Softw.
  Engg.}, vol.~21, no.~1, pp. 17--42, Feb. 2016.

\bibitem{Rayner1998}
K.~Rayner, ``Eye movements in reading and information processing: 20 years of
  research,'' vol. 124, no.~3, pp. 372--422, 00 1998.

\bibitem{Olsson2007}
P.~Olsson, ``Real-time and offline filters for eye tracking,'' p.~42, 2007.

\bibitem{Collard2011}
M.~L. Collard, M.~J. Decker, and J.~I. Maletic, ``Lightweight transformation
  and fact extraction with the srcml toolkit,'' in \emph{2011 IEEE 11th
  International Working Conference on Source Code Analysis and Manipulation},
  2011, pp. 173--184.

\bibitem{Liblit2006}
B.~Liblit, A.~Begel, and E.~Sweetser, ``Cognitive perspectives on the role of
  naming in computer programs,'' in \emph{Proceedings of the 18th Annual
  Psychology of Programming Workshop}, ser. PPIG '06, Brighton, United Kingdom,
  2006.

\bibitem{siegel1956nonparametric}
S.~Siegel, \emph{Nonparametric statistics for the behavioral sciences}, ser.
  McGraw-Hill series in psychology.

\bibitem{Abdi2007}
\BIBentryALTinterwordspacing
H.~Abdi. (2007) The kendall rank correlation coefficient. [Online]. Available:
  \url{http://www.utdallas.edu/~herve/Abdi- KendallCorrelation2007-pretty.pdf}
\BIBentrySTDinterwordspacing

\end{thebibliography}

\end{document}